\documentclass[aps,pra,twocolumn,
  superscriptaddress, longbibliography,
  nofootinbib]{revtex4-2}

\usepackage{amsmath,amssymb,mathtools}
\usepackage{bm}
\usepackage{bbm}
\usepackage{graphicx}
\usepackage{microtype}
\usepackage[colorlinks=true,citecolor=blue,linkcolor=blue,urlcolor=blue]{hyperref}

\newcommand{\e}{\mathrm{e}}
\newcommand{\Tr}{\operatorname{Tr}}
\newcommand{\sech}{\operatorname{sech}}
\newcommand{\cO}{\mathcal O}
\newcommand{\cL}{\mathcal L}

\newcommand{\cl}{\mathrm{cl}}
\newcommand{\nc}{\mathrm{nc}}
\newcommand{\LambdaK}{\Lambda_{\mathrm K}}

\begin{document}

\title{Hall's Exact Uncertainty Relation and the Geometry of Krylov-Space Operator Dynamics}

\author{Mohsen Alishahiha}
\email{alishah@ipm.ir}
\affiliation{School of Quantum Physics and Matter, Institute for Research in Fundamental Sciences (IPM), P.O. Box 19395-5531, Tehran, Iran}

\author{Souvik Banerjee}
\email{souvik.banerjee@christuniversity.in}
\affiliation{Department of Physics and Electronics, Christ University, Bengaluru 560029, India}
\affiliation{Institut f{\"u}r Theoretische Physik und Astrophysik, Julius-Maximilians-Universit{\"a}t W{\"u}rzburg, Am Hubland, 97074 W{\"u}rzburg, Germany}

\begin{abstract}
Krylov representations of operator dynamics encode two inequivalent motions:
probability transport along the discrete Krylov chain and motion of the
normalized amplitude vector on the associated sphere. We lift Hall's exact
uncertainty relation to Liouville space and show that, for the fixed
Krylov-basis measurement, its classical Fisher information locally attains the
quantum Fisher information. The amplitude vector therefore follows an exact
constant-speed spherical trajectory. We construct its moving frame,
identifying the first Lanczos coefficient with the speed and the higher
coefficients with successive bending. Geodesic curvature controls the leading
departure from Mandelstam-Tamm saturation. Separately, we compare cumulative
and pointwise bounds on Krylov-index transport, which depends on the full
Lanczos sequence.
\end{abstract}

\maketitle

%%%%%%%%%%%%%%%%%%%%%%%%%%%%%%%%%%%%%%%%%%%%%%%%%%%%%%%%%%%%%%%%%%%%%%%%%%%%%%%

\section{Introduction}
\label{sec:introduction}

The uncertainty principle is a defining feature of quantum mechanics, limiting the precision with which certain pairs of observables can be simultaneously measured. Beyond the canonical position–momentum relation, the energy–time uncertainty relations~\cite{Mandelstam, Margolus:1997ih} imply fundamental limits on how rapidly a quantum state may evolve, known as quantum speed limits (QSLs). Such limits play a central role in quantum control, quantum information processing, and quantum computation.

Most uncertainty relations take the form of \emph{inequalities}, and therefore yield lower bounds on the time required for a given evolution. In contrast, Hall derived a state-dependent \emph{exact} uncertainty relation for two observables $A$ and $B$ in a pure state $\rho = |\psi\rangle\langle\psi|$~\cite{Hall:2001zz},
\begin{equation}
\delta_B A\,\Delta B_{\nc}=\frac{\hbar}{2},
\label{eq:hall-intro}
\end{equation}
where $B_{\nc}=B-B_{\cl}$ is the part of $B$ that is nonclassical relative to
a projective measurement of $A$, and $(\delta_B A)^{-2}$ is the classical
Fisher information carried by the corresponding measurement probabilities
under the flow generated by $B$. The classical--nonclassical decomposition is
therefore measurement relative rather than basis independent. Hall's equality
has subsequently been used to formulate exact path-length relations for
unitary and Liouville-space dynamics \cite{Pati:2023seq,Srivastav:2024apk}.
Definitions and technical derivations used below are provided in the appendixes.

Closely related questions arise for Heisenberg operators. One may ask how rapidly an operator state moves in operator Hilbert space, how rapidly its overlap with the initial operator changes, or how rapidly its weight spreads over a chosen basis. These questions involve different observables and, as will be central below, need not use the same metric or the same notion of speed.

Krylov space provides a natural representation of operator dynamics. Starting
from a normalized seed operator, the Lanczos construction maps the Liouvillian
to a nearest-neighbor chain with hopping amplitudes given by the Lanczos
coefficients $\{b_1,b_2,\ldots\}$
\cite{lanczos1950,viswanath2008recursion}. This representation has become a
standard framework for operator growth and Krylov complexity
\cite{Parker:2018yvk,Barbon:2019wsy}. It has also been used to derive
uncertainty bounds on Krylov-complexity growth \cite{Hornedal:2022pkc}, QSLs
for operator flows \cite{Carabba:2022operatorflow}, geometric operator QSLs
\cite{Hornedal:2023xpa}, and a hierarchy of Krylov-space speed and propagation
limits based on nonescape probabilities \cite{TakahashiDelCampo:2025}.
Related recent work compared Krylov state complexity with
information-geometric measures in qubit dynamics and developed a
resolvent-based Krylov formulation of quantum Fisher information
\cite{CafaroClementsAnuboyina2026,AlishahihaTabeshVasli2026}. These studies
address state-complexity geometry and QFI computation, respectively, whereas
the present work identifies a fixed Krylov projective measurement that attains
the time-translation QFI of an evolving operator state.

The present work is organized around two inequivalent motions encoded by the
same amplitudes. First, the probabilities $p_n(t)$ are redistributed over the
discrete Krylov index $n$. This is a transport problem measured with the graph
distance $|m-n|$ and characterized by quantities such as Krylov complexity,
nonescape probabilities, and probability tails. Its behavior depends on the
full Lanczos sequence. Second, the complete real amplitude vector $\Phi(t)$
moves on the unit sphere implied by normalization. This is a continuous motion
in operator Hilbert space, measured by the metric inherited from the ambient
Euclidean space. The two notions of distance are inequivalent: A large index
displacement need not imply a large spherical distance.

In contrast to the nonescape-based propagation hierarchy of
Ref.~\cite{TakahashiDelCampo:2025}, our starting point is Hall's exact
uncertainty relation applied to the normalized operator state in Liouville
space. Hall's theorem itself 
requires no algebraic modification: A normalized operator is a 
pure state in Hilbert-Schmidt space, the Liouvillian is its 
Hermitian generator, and a projective operator-space observable may 
be used as $A$. The nontrivial specialization is obtained when that
observable has the Krylov vectors as its eigenbasis. We show that the 
classical component of the Liouvillian then vanishes relative to this 
fixed measurement and that the resulting classical Fisher information
attains the quantum Fisher information at every time,
\begin{equation}
F_K=F_Q=4b_1^2.
\label{eq:intro-FI}
\end{equation}
Equivalently, the amplitude vector moves on the Krylov sphere with the
exact constant metric speed $b_1$.

A central observation is the separation between the exact 
path-length equality
and the usual endpoint QSL. Hall's relation determines the length of the
actual spherical path. The Mandelstam-Tamm bound follows only 
after the length is replaced by the shortest endpoint distance. 
The actual path is generically nongeodesic; nongeodesicity is 
therefore not the condition for validity of the
Mandelstam-Tamm inequality, but the geometric origin of 
its nonsaturation. We construct an exact spherical moving frame 
in which $b_1$ fixes the speed,
$b_2/b_1$ is the geodesic curvature, and the higher Lanczos coefficients
determine successive bending into new Krylov directions. This construction is
compatible with earlier moment-based Frenet geometry in projective Hilbert
space \cite{AlsingCafaro:2024}; its additional content is the complete
Krylov-resolved hierarchy of connection coefficients. It is also distinct
from phase-space geometries of Krylov complexity \cite{Caputa:2021sib}.

Finally, we keep index-space propagation logically separate from the Hall
geometry. We compare this directly with the nonescape hierarchy of
Ref.~\cite{TakahashiDelCampo:2025} and retain a complementary
individual-amplitude estimate: If the Lanczos Jacobi matrix has a
finite global row-sum bound, each $\phi_n(t)$ obeys a factorial 
tail bound. This estimate is related to standard bounds for Jacobi 
operators \cite{Teschl2000} but is useful
because it controls individual Krylov amplitudes rather than only cumulative
probabilities. For unbounded Lanczos coefficients, no finite global index
velocity follows from this estimate. Under a slowly varying, predominantly
right-moving envelope approximation, the leading continuum current gives
$\dot n\simeq2b(n)$; in particular, $b_n\sim\alpha n$ produces the scale
$\e^{2\alpha t}$, consistent with exactly solvable Krylov chains.

This paper is organized as follows. Section~\ref{sec:hall-operator} develops
Hall's exact relation for operator dynamics and its Fisher-information
interpretation. Section~\ref{sec:sphere} constructs the geometry of the
Krylov-sphere trajectory and relates geodesicity to saturation of the
Mandelstam--Tamm bound. Section~\ref{sec:index} treats propagation in the
Krylov index, including nonescape bounds, the individual-amplitude tail
estimate, and the continuum factor of 2 for growing Lanczos coefficients.
Section~\ref{sec:conclusion} concludes. Technical derivations
and examples are collected in the appendixes. Throughout the paper we have set $\hbar=1$.

%%%%%%%%%%%%%%%%%%%%%%%%%%%%%%%%%%%%%%%%%%%%%%%%%%%%%%%%%%%%%%%%%%%%%%%%%%%%%%%

\section{Hall's Exact Relation for Operator Dynamics}
\label{sec:hall-operator}

The Introduction distinguished two inequivalent motions encoded by the same
Krylov amplitudes: discrete probability transport along the Krylov chain and
continuous motion of the normalized amplitude vector on the Krylov sphere.
This section develops the latter. We show that Hall's exact uncertainty
relation, lifted to Liouville space, provides a measurement-theoretic
interpretation of the spherical motion and identifies its constant speed with
the first Lanczos coefficient $b_1$.

\subsection{Krylov representation and constant spherical speed}
\label{subsec:krylov}

Consider a closed quantum system with a time-independent Hermitian Hamiltonian $H$ and a nonzero Hermitian seed operator $\mathcal{O}(0)$.~\footnote{In what follows we consider operator growth, 
though the construction generalizes to state spreading~\cite{Balasubramanian:2022tpr, Alishahiha:2022anw,
Caputa:2024vrn, Alishahiha:2022nhe}.} We use
\begin{equation}
\mathcal{O}(t)=\mathrm{e}^{-iHt}\mathcal{O}(0)\mathrm{e}^{iHt},
\qquad
\mathcal{L}\mathcal{O}=[H,\mathcal{O}],
\label{eq:operator-flow}
\end{equation}
so that $i\partial_t|\mathcal{O}(t))=\mathcal{L}|\mathcal{O}(t))$.

The Operator Hilbert space is equipped with the Hilbert-Schmidt inner product $(A|B)=\Tr(A^\dagger B)$, and the seed is normalized, $(\mathcal{O}(0)|\mathcal{O}(0))=1$.

The Lanczos procedure generates an orthonormal Krylov basis $\{|K_n)\}$ satisfying
\begin{equation}
|K_0)=|\mathcal{O}(0)),
\qquad
\mathcal{L}|K_n)=b_{n+1}|K_{n+1})+b_n|K_{n-1}),
\label{eq:lanczos}
\end{equation}
with $b_0=0$ and $b_n\geq0$. For the Hermitian seed and commutator Liouvillian considered here, the diagonal Lanczos coefficients vanish. The normalized operator state can then be expanded as
\begin{equation}
|\mathcal{O}(t))=\sum_{n=0}^{D_{\mathcal{O}}-1}(-i)^n\phi_n(t)|K_n),
\qquad
\sum_{n=0}^{D_{\mathcal{O}}-1}\phi_n^2(t)=1,
\label{eq:expansion}
\end{equation}
where the amplitudes are real, $\phi_n(0)=\delta_{n0}$, and
\begin{equation}
\dot\phi_n=b_n\phi_{n-1}-b_{n+1}\phi_{n+1}.
\label{eq:amplitude-recursion}
\end{equation}
Here $D_{\mathcal{O}}$ is the dimension of Krylov space
associated with the seed operator $\cal O$.
We assume $b_1>0$ and, in an infinite-dimensional Krylov space, that the evolving state lies in the domain of $\mathcal{L}^2$.

The amplitude vector
\begin{equation}
\Phi(t)=\bigl(\phi_0(t),\phi_1(t),\ldots\bigr)^T
\end{equation}
lies on the unit sphere in the Krylov representation. Its norm speed obeys
\begin{align}
\|\dot\Phi(t)\|^2
&=\sum_n\dot\phi_n^2(t)
=(\mathcal{O}(t)|\mathcal{L}^2|\mathcal{O}(t))
\nonumber\\
&=(K_0|\mathcal{L}^2|K_0)=b_1^2,
\label{eq:constant-speed}
\end{align}
where time independence of $\mathcal{L}$ was used in the second equality. This identifies a velocity
\begin{equation}
v_{\mathrm{sph}}\equiv\|\dot\Phi(t)\|=b_1,
\label{eq:vsph}
\end{equation}
which is strictly constant in time along the trajectory on the sphere.

Although the constancy of $v_{\mathrm{sph}}$ follows directly from conservation of the Liouvillian variance, its physical interpretation is not evident from the Krylov recursion alone. The next section shows that Hall's exact uncertainty relation identifies this constant speed as the Fisher-optimal statistical speed associated with a fixed Krylov measurement. This distinction—between a kinematic consequence and its operational interpretation—is central to what follows.

\subsection{Liouville-space lift of Hall's equality}
\label{subsec:hall-lift}

To lift Hall's uncertainty relation to operator dynamics, we observe that a normalized operator defines a pure state in operator Hilbert space,
\begin{equation}
\varrho(t)=|\mathcal{O}(t))(\mathcal{O}(t)|.
\end{equation}
Equivalently, under vectorization this is a pure state on a doubled Hilbert space, with the Liouvillian represented by $H\otimes\mathbbm 1-\mathbbm 1\otimes H^T$. We define the nondegenerate operator-space observable
\begin{equation}
A_K=\sum_n a_n|K_n)(K_n|,
\qquad a_n\neq a_m\quad(n\neq m),
\label{eq:AK}
\end{equation}
whose eigenbasis is precisely the Krylov basis, and choose $B=\mathcal{L}$ in Eq.~\eqref{eq:hall-intro}. The outcome probabilities for a projective measurement of $A_K$ are
\begin{equation}
p_n(t)=(K_n|\varrho(t)|K_n)=\phi_n^2(t).
\label{eq:probabilities}
\end{equation}

Relative to this Krylov-basis measurement, Hall's classical component of the Liouvillian vanishes. Indeed, with $c_n=(-i)^n\phi_n$ and $i\dot c=\mathcal{L}c$,
\begin{equation}
(K_n|\tfrac12\{\mathcal{L},\varrho\}|K_n)
=\operatorname{Re}\!\left[c_n^*(\mathcal{L}c)_n\right]=0.
\end{equation}
Consequently,
\begin{equation}
\mathcal{L}_{\mathrm{cl}}^{(K)}=0,
\qquad
\mathcal{L}_{\mathrm{nc}}^{(K)}=\mathcal{L}.
\label{eq:Lclzero}
\end{equation}
The superscript $(K)$ is essential: This statement is relative to the Krylov projective measurement and is not a basis-independent property of $\mathcal{L}$. The decomposition into classical and nonclassical parts is therefore measurement relative, exactly as in Hall's original formulation.

The Liouville-space lift itself is formally straightforward. The nontrivial content is the specialization to the Krylov measurement, for which the classical component vanishes identically and Hall's equality acquires the geometric interpretation developed below. The classical Fisher information associated with this measurement is
\begin{align}
F_K(t)
&=\sum_n\frac{\dot p_n^2(t)}{p_n(t)}
=4\sum_n\dot\phi_n^2(t)
\nonumber\\
&=4b_1^2,
\label{eq:FK}
\end{align}
where terms at isolated zeros of $p_n$ are defined by continuous extension. For the pure operator state, the quantum Fisher information for time translations generated by $\mathcal{L}$ is~\cite{BraunsteinCaves:1994}
\begin{equation}
F_Q=4\bigl[(\mathcal{O}|\mathcal{L}^2|\mathcal{O})-(\mathcal{O}|\mathcal{L}|\mathcal{O})^2\bigr]=4b_1^2.
\label{eq:FQ}
\end{equation}
Therefore,
\begin{equation}
F_K(t)=F_Q=4b_1^2.
\label{eq:FI-equality}
\end{equation}

This is the principal conceptual result of the present section: the fixed Krylov measurement extracts all time-translation Fisher information carried by the pure operator state. Thus no alternative measurement can reveal a greater statistical speed for the operator dynamics than the fixed Krylov measurement itself. This Fisher optimality, rather than the formal possibility of viewing an operator as a Liouville-space state, is the special content of the Krylov construction. The result is therefore not a generic consequence of vectorizing operators into Liouville space; it relies on the dynamical adaptation of the Krylov basis, which renders the Liouvillian purely nonclassical relative to the associated projective measurement.

The operational statement has an important qualification. The Krylov
projectors are fixed once the Hamiltonian and seed operator are specified, but
their realization may require a collective or nonlocal measurement on the
vectorized operator state and need not correspond to a simple measurement in
the original many-body Hilbert space.

\subsection{Exact Hall path length}
\label{subsec:hall-length}

We adopt the Fisher--Rao convention
\begin{equation}
d\ell_{\mathrm{FR}}^2=\frac14\sum_n\frac{dp_n^2}{p_n}.
\end{equation}
Equation~\eqref{eq:probabilities} gives
\begin{equation}
d\ell_{\mathrm{FR}}^2=\sum_n d\phi_n^2\equiv d\ell_{\mathrm{sph}}^2.
\label{eq:fisher-sphere}
\end{equation}
The signed amplitudes provide a smooth unfolding of the usual square-root probability coordinates. Equations~\eqref{eq:vsph} and \eqref{eq:FI-equality} imply
\begin{equation}
\ell_{\mathrm{sph}}(0,T)=\int_0^Tdt\,\|\dot\Phi(t)\|=b_1T.
\label{eq:exact-length}
\end{equation}
Since $\Delta\mathcal{L}_{\mathrm{nc}}^{(K)}=b_1$, Hall's equality takes the exact speed-relation form
\begin{equation}
T=\frac{\ell_{\mathrm{sph}}(0,T)}{\Delta\mathcal{L}_{\mathrm{nc}}^{(K)}}
=\frac{\ell_{\mathrm{sph}}(0,T)}{b_1}.
\label{eq:hall-speed-relation}
\end{equation}

This equality involves the length of the actual trajectory on the Krylov sphere. The conventional Mandelstam-Tamm endpoint QSL follows only after  this actual path length is replaced by the shortest geodesic distance between the endpoints. The distinction between the exact Hall relation and the inequality is therefore purely geometric: The former fixes the length of the path taken, while the latter bounds the time by the minimal possible distance. This explains why Hall's equality and the Mandelstam-Tamm bound are not competing speed limits: They quantify different geometric objects, namely the actual trajectory and the shortest connecting geodesic. We shall explore the consequences of this nongeodesicity in the next section.

%%%%%%%%%%%%%%%%%%%%%%%%%%%%%%%%%%%%%%%%%%%%%%%%%%%%%%%%%%%%%%%%%%%%%%%%%%%%%%%

\section{Geometry of the Krylov-Sphere Trajectory}
\label{sec:sphere}

The previous section established that Hall's uncertainty relation endows the
Krylov-sphere trajectory with a constant locally Fisher-optimal speed $b_1$.
We now turn to the shape of this trajectory. While $b_1$ fixes the speed, the
higher Lanczos coefficients determine how the path bends on the sphere.

\subsection{Spherical moving frame}
\label{subsec:moving-frame}

Equation~\eqref{eq:amplitude-recursion} can be written as
$\dot\Phi=\mathsf A\Phi$, where $\mathsf A$ is a real skew-symmetric
tridiagonal matrix. Let $\bm e_n$ denote the fixed coordinate vectors and
define
\begin{equation}
\bm E_n(t)=\e^{t\mathsf A}\bm e_n.
\label{eq:frame-definition}
\end{equation}
$\bm E_n$ form an orthonormal moving frame, with $\bm E_0=\Phi$. Using
the arc-length parameter $s=b_1t$,we obtain
\begin{subequations}
\label{eq:spherical-frame}
\begin{align}
\frac{d\bm E_0}{ds}&=\bm E_1,
\label{eq:frame0}\\
\frac{d\bm E_1}{ds}&=-\bm E_0+\kappa_1^{(S)}\bm E_2,
\label{eq:frame1}\\
\frac{d\bm E_n}{ds}
&=-\kappa_{n-1}^{(S)}\bm E_{n-1}
+\kappa_n^{(S)}\bm E_{n+1},\qquad n\geq2,
\label{eq:framen}
\end{align}
\end{subequations}
where
\begin{equation}
\kappa_n^{(S)}=\frac{b_{n+1}}{b_1},
\qquad n\geq1.
\label{eq:spherical-curvatures}
\end{equation}
This moving frame provides a complete hierarchical classification of the trajectory's shape. The first coefficient $b_1$ fixes the parametrization speed, while the normalized higher Lanczos coefficients determine successive bending into new Krylov directions. In particular,
\begin{equation}
\kappa_g=\frac{b_2}{b_1}
\label{eq:geodesic-curvature}
\end{equation}
is the geodesic curvature of the spherical path. The 
first ambient Euclidean curvature and torsion are
\begin{equation}
\kappa_E=\sqrt{1+\frac{b_2^2}{b_1^2}},
\qquad
|\tau_E|=\frac{b_2b_3}{b_1\sqrt{b_1^2+b_2^2}}.
\label{eq:curvature-torsion}
\end{equation}
The spherical hierarchy in Eq.~\eqref{eq:spherical-frame} is the
dimension-independent statement; detailed derivations are given in
the appendixes.

The relation to earlier projective-Hilbert-space geometry can be made precise.
Because
\begin{equation}
(\cO|\dot{\cO})=-i(\cO|\cL|\cO)=0,
\end{equation}
the normalized Liouville-space trajectory is a horizontal lift of its ray in
projective operator Hilbert space. Consequently, the spherical metric used
here agrees locally with the Fubini--Study metric along the trajectory. The
first intrinsic curvature may equivalently be written as
\begin{equation}
\kappa_g^2
=\frac{\langle\cL^4\rangle-\langle\cL^2\rangle^2}
{\langle\cL^2\rangle^2}
=\frac{b_2^2}{b_1^2},
\label{eq:moment-curvature}
\end{equation}
where $\langle\cL^m\rangle=(\cO|\cL^m|\cO)$. Equation
\eqref{eq:moment-curvature} is compatible with earlier moment-based
formulations of quantum-trajectory curvature \cite{AlsingCafaro:2024}. The
additional content of the present construction is the complete
Krylov-resolved moving-frame hierarchy, which identifies every successive
connection coefficient directly with a normalized Lanczos coefficient.

\subsection{Nongeodesicity and the Mandelstam-Tamm bound}
\label{subsec:MT}

For a unit-speed curve on the unit sphere, the covariant acceleration is the tangential part of $d\bm E_1/ds$—that is, the component of the ambient acceleration orthogonal to the velocity, which measures how the velocity direction changes within the sphere. If we denote this covariant derivative along the curve by $D/ds$, Eq.~\eqref{eq:frame1} gives
\begin{equation}
\frac{D\bm E_1}{ds}=\frac{b_2}{b_1}\bm E_2.
\label{eq:covariant-acceleration}
\end{equation}
Hence the Krylov-sphere trajectory is geodesic if and only if
\begin{equation}
b_2=0.
\label{eq:geodesic-condition}
\end{equation}
In the Lanczos construction, this condition terminates the chain after
$|K_1)$ and confines the dynamics to a great circle. Generic chains with
$b_2>0$ are nongeodesic. Explicit geodesic and nongeodesic examples are given
in Appendix \ref{app:qubits}.

The conventional projective Mandelstam-Tamm distance involves
$\arccos|\phi_0(t)|$. On the initial branch considered below, the bound itself
ensures $\phi_0(t)\geq0$, so the projective and signed-sphere expressions
coincide. The spherical distance from the initial point is therefore
\begin{equation}
\theta(t)=d_{\mathrm{sph}}\bigl(\Phi(0),\Phi(t)\bigr)
=\arccos\phi_0(t).
\label{eq:spherical-distance}
\end{equation}
Since the shortest distance on the sphere cannot exceed the length of the actual path,
\begin{equation}
\theta(t)\leq\ell_{\mathrm{sph}}(0,t)=b_1t.
\label{eq:MT-angle}
\end{equation}
For $0\leq b_1t\leq\pi/2$, this becomes
\begin{equation}
\phi_0(t)\geq\cos(b_1t).
\label{eq:return-bound}
\end{equation}
This is the Mandelstam-Tamm survival-amplitude bound in Krylov coordinates,
not a new index-space propagation bound. It holds whether or not the trajectory
is geodesic. Geodesicity determines saturation: Equality over a finite initial
interval occurs if and only if $b_2=0$.

The leading gap from saturation makes this statement quantitative. The short-time expansion gives
\begin{equation}
\phi_0(t)-\cos(b_1t)
=\frac{b_1^2b_2^2}{24}t^4+O(t^6)
=\frac{b_1^4\kappa_g^2}{24}t^4+O(t^6).
\label{eq:MT-gap}
\end{equation}
Thus $b_2/b_1$ has a direct dynamical meaning: it controls the first departure from a Mandelstam--Tamm-saturating great-circle trajectory.
Higher Lanczos coefficients govern successive 
local bending invariants of the trajectory while leaving the constant Hall speed $b_1$ unchanged. 
The spherical moving-frame construction therefore separates three inequivalent aspects of operator dynamics that are often conflated:
the statistical speed of evolution $b_1$, the shape of the trajectory (the higher normalized Lanczos coefficients), and the transport of probability along the Krylov chain.
They are complementary rather than interchangeable descriptions of operator dynamics.

%%%%%%%%%%%%%%%%%%%%%%%%%%%%%%%%%%%%%%%%%%%%%%%%%%%%%%%%%%%%%%%%%%%%%%%%%%%%%%%
\section{Propagation in Krylov-Index Space}
\label{sec:index}

We now turn to the second of the two inequivalent motions introduced in Sec.~\ref{sec:introduction}: the transport of probability along the discrete Krylov chain. While the previous sections established that the full amplitude vector moves on the Krylov sphere with the exact constant Hall speed $b_1$, the same amplitudes also define a probability distribution $p_n=\phi_n^2$ on the discrete index $n$. The graph metric of this chain is inequivalent to the spherical metric. Let $\bm e_n$ denote the canonical basis vector in amplitude space—the state perfectly localized at a single Krylov site $n$, so that $\phi_n=1$ and all other amplitudes vanish. For two distinct localized states $\bm e_m$ and $\bm e_n$ with $m\neq n$, their inner product is zero, and hence the spherical distance between them is
\begin{equation}
d_{\mathrm{sph}}(\bm e_m,\bm e_n)=\frac{\pi}{2}\quad(m\neq n).
\label{eq:sphere-distance-localized}
\end{equation}
By contrast, the graph distance along the Krylov chain is simply the index difference,
\begin{equation}
d_{\mathrm{idx}}(m,n)=|m-n|.
\label{eq:idx-distance-localized}
\end{equation}
Consequently, the Hall path length cannot by itself determine how far probability has propagated in the index. A large index distance need not imply a large spherical distance, and vice versa. This section reviews the cumulative propagation hierarchy of Ref.~\cite{TakahashiDelCampo:2025} and develops complementary bounds for individual amplitudes and continuum transport.

\subsection{Nonescape and individual-amplitude bounds}
\label{subsec:bounds}

A useful cumulative diagnostic is the nonescape probability
\begin{equation}
P_N(t)=\sum_{n=0}^{N}p_n(t),
\qquad
\Theta_N(t)=\arccos\sqrt{P_N(t)}.
\label{eq:nonescape}
\end{equation}
For the time-independent off-diagonal Krylov chain, the hierarchy of Ref.~\cite{TakahashiDelCampo:2025} implies
\begin{equation}
|\dot\Theta_N(t)|
\leq b_{N+1}\sin \Theta_{N-1}(t)\leq b_{N+1}\Theta_{N-1}(t),
\label{eq:nonescape-hierarchy}
\end{equation}
and hence
\begin{equation}
\Theta_N(t)
\leq
\frac{b_1b_2\cdots b_{N+1}}{(N+1)!}|t|^{N+1}.
\label{eq:local-product}
\end{equation}
At $N=0$, this recovers the Mandelstam-Tamm bound for the survival amplitude; for $N>0$, it constrains escape beyond the first $N+1$ sites.

The hierarchy of Ref.~\cite{TakahashiDelCampo:2025} constrains cumulative escape probabilities. We now consider a complementary pointwise question. Suppose that the Lanczos
matrix has the finite global row-sum scale
\begin{equation}
\LambdaK\equiv\sup_{n\geq0}(b_n+b_{n+1})<\infty.
\label{eq:LambdaK}
\end{equation}
Tridiagonality gives the rigorous estimate
\begin{equation}
|\phi_n(t)|
\leq
\sum_{m=n}^{\infty}\frac{(\LambdaK|t|)^m}{m!}
\leq
\e^{\LambdaK|t|}\frac{(\LambdaK|t|)^n}{n!}.
\label{eq:amplitude-tail}
\end{equation}
Equivalently,
\begin{equation}
|\phi_n(t)|
\lesssim
\exp\!\left[
-n\ln\!\left(\frac{n}{\LambdaK|t|}\right)
+n+\LambdaK|t|
\right],
\label{eq:decay}
\end{equation}
which is exponentially small when $n\gg\LambdaK|t|$. Whereas the hierarchy
of Ref.~\cite{TakahashiDelCampo:2025} constrains cumulative escape,
Eq.~\eqref{eq:amplitude-tail} controls each individual amplitude. It is a
uniform pointwise bound for evolution launched from the initial Krylov vector
$K_0$, not a seed-independent statement.

The proof is given in Appendix \ref{app:tail}. The
estimate is related to standard banded-matrix and Jacobi-operator bounds
\cite{Teschl2000}; we do not identify it as a consequence of Hall's equality
or claim it as a new universal Lieb-Robinson bound. 
The local-product bound \eqref{eq:local-product} and the global bound \eqref{eq:amplitude-tail} are complementary. The former retains detailed local information about the sequence $b_1,\ldots,b_{N+1}$ and is generally sharper for small $N$. The latter is simpler and uniform in $n$, but applies only when a finite global row-sum bound exists. The two bounds therefore answer different questions: cumulative escape versus pointwise amplitude suppression.
 For a uniform semi-infinite
chain $b_n=b$, the exact solution has ballistic scale $n\sim2bt$, consistent
with $\LambdaK=2b$.

\subsection{Growing Lanczos coefficients and the factor of 2}
\label{subsec:factor-two}

If $b_n$ is unbounded, $\LambdaK$ diverges and
Eq.~\eqref{eq:amplitude-tail} does not furnish a finite global index scale. A
slowly varying, predominantly right-moving envelope nevertheless admits a
continuum approximation. Expanding Eq.~\eqref{eq:amplitude-recursion} at large
$n$ gives, to leading order,
\begin{equation}
\partial_t p(n,t)
+\partial_n\!\left[2b(n)p(n,t)\right]\simeq0,
\label{eq:continuity}
\end{equation}
whose characteristics obey
\begin{equation}
\dot n_{\mathrm{char}}(t)
\simeq2b\bigl(n_{\mathrm{char}}(t)\bigr).
\label{eq:characteristics}
\end{equation}
Equation~\eqref{eq:characteristics} is an asymptotic envelope estimate, not a
universal rigorous front bound. Under its stated assumptions, the leading
advection velocity is $2b(n)$, reproducing the scaling of the exactly solvable
chains discussed in Appendix \ref{app:growing}.

Equation~\eqref{eq:characteristics} yields the characteristic scalings
\begin{equation}
\begin{array}{rcl}
b(n)=b &\Longrightarrow& n_{\mathrm{typ}}(t)\sim2bt,\\[1mm]
b(n)\sim\alpha\sqrt n &\Longrightarrow& n_{\mathrm{typ}}(t)\sim(\alpha t)^2,\\[1mm]
b(n)\sim\alpha n &\Longrightarrow& n_{\mathrm{typ}}(t)\propto\mathrm{e}^{2\alpha t}.
\end{array}
\label{eq:scalings}
\end{equation}

The constant, square-root, and linear cases therefore give ballistic,
quadratic, and exponential index scales, respectively. In particular, the
factor of 2 in the exponent for asymptotically linear coefficients agrees
with the exact negative-binomial Krylov chain.

These results make the two-motion distinction explicit. Index-space transport may be ballistic, diffusive-like, polynomial, or exponential depending on the Lanczos sequence, whereas the Hall trajectory always accumulates Fisher-Rao length at the constant rate $b_1$. The two statements concern inequivalent metrics; the spherical speed is not a Lieb-Robinson velocity in the index. This is precisely the distinction advertised in the Introduction: The Hall geometry governs the continuous motion of the full amplitude vector on the Krylov sphere, while index-space propagation—whether characterized by nonescape probabilities, individual amplitude tails, or the growth of Krylov complexity—probes a different notion of distance and is sensitive to the full Lanczos sequence rather than just $b_1$.

%%%%%%%%%%%%%%%%%%%%%%%%%%%%%%%%%%%%%%%%%%%%%%%%%%%%%%%%%%%%%%%%%%%%%%%%%%%%%%%

\section{Conclusion}
\label{sec:conclusion}
Krylov operator dynamics encodes two inequivalent notions of motion: continuous
motion of the normalized amplitude vector on the Krylov sphere and transport
of probability along the discrete Krylov chain. These motions are governed by
different metrics and quantify different physical aspects of operator
evolution.

By regarding a normalized operator as a pure state in Liouville space, we  formulated Hall's exact uncertainty relation for Krylov dynamics. The Liouville-space lift itself is formally straightforward; the special content of the Krylov construction is that the fixed projective measurement in the Krylov basis makes the classical part of the Liouvillian vanish and attains the quantum Fisher information at every time,
\begin{equation}
F_K=F_Q=4b_1^2.
\label{eq:conclusion-FI}
\end{equation}
This identifies $b_1$ as the exact Fisher-optimal statistical speed of the full amplitude vector on the Krylov sphere. The Hall equality therefore supplies an operational, measurement-theoretic interpretation of the constant spherical speed that is not evident from the Krylov recursion alone.

Hall's equality fixes the length of the actual spherical trajectory, whereas
the Mandelstam-Tamm inequality follows after  that length is replaced by the
shortest endpoint distance. Generic trajectories are nongeodesic, and this is
the geometric origin of nonsaturation. The moving frame shows that $b_2/b_1$
is the geodesic curvature and that the higher normalized Lanczos coefficients
form a hierarchy of successive bending coefficients. In particular, $b_2$
controls the leading $t^4$ departure from Mandelstam-Tamm saturation.

Propagation in the discrete index is a separate problem. The established
nonescape hierarchy constrains cumulative escape, while the row-sum estimate
provides a complementary pointwise amplitude bound for bounded Lanczos chains.
For growing coefficients, the continuum envelope equation gives the
characteristic law $\dot n\simeq2b(n)$ under its stated assumptions, yielding
ballistic, quadratic, or exponential scales according to the asymptotic growth
of $b_n$. None of these index-space laws follows from the spherical path length
alone.

Keeping the two geometries distinct clarifies both the physical content and
the limitations of geometric speed statements in Krylov space and shows how
they describe complementary aspects of the same operator evolution.

\section*{Acknowledgments}
We thank Mohammad Reza Tanhayi and Mohammad Javad Vasli for insightful discussions on complexity, and Arun K. Pati for comments on an earlier version of this article. The work of M.A. is supported by the Iran National Science Foundation (INSF) under Project No.~4023620. This work is dedicated to Sudipta Mukherji on the occasion of his 60th birthday.

%%%%%%%%%%%%%%%%%%%%%%%%%%%%%%%%%%%%%%%%%%%%%%%%%%%%%%%%%%%%%%%%%%%%%%%%%%%%%%%

\onecolumngrid
\appendix
%\section*{Supplemental Material for ``Hall's Exact Uncertainty Relation and the Geometry of Krylov-Space Operator Dynamics''}
%\makeatletter
%\@removefromreset{equation}{section}
%\makeatother
%\setcounter{equation}{0}
%\renewcommand{\theequation}{S\arabic{equation}}
%\setcounter{section}{0}
%\renewcommand{\thesection}{S\Roman{section}}

%%%%%%%%%%%%%%%%%%%%%%%%%%%%%%%%%%%%%%%%%%%%%%%%%%%%%%%%%%%%%%%%%%%%%%%%%%%%%%%
%%%%%%%%%%%%%%%%%%%%%%%%%%%%%%%%%%%%%%%%%%%%%%%%%%%%%

%%%%%%%%%%%%%%%%%%%%%%%%%%%%%%%%%%%%%%%%%%%%%%%%%%%%%%%%%%%%%%%%%%%%%%%%%%%%%%%

\section{Hall's relation and its Liouville-space realization}
\label{supp:hall}

Let $A$ and $B$ be observables and let
$\rho=|\psi\rangle\langle\psi|$ be a pure state. If
$\{|a_k\rangle\}$ is a nondegenerate eigenbasis of $A$, Hall's exact
uncertainty relation reads \cite{Hall:2001zz}
\begin{equation}
\delta_B A\,\Delta B_{\nc}=\frac12,
\label{eq:S-hall}
\end{equation}
where
\begin{equation}
(\delta_B A)^{-2}
=\sum_k
\frac{\langle a_k|i[B,\rho]|a_k\rangle^2}
{\langle a_k|\rho|a_k\rangle}
\label{eq:S-hall-FI}
\end{equation}
and
\begin{equation}
B_{\cl}
=\sum_k
\frac{\langle a_k|\tfrac12\{B,\rho\}|a_k\rangle}
{\langle a_k|\rho|a_k\rangle}
|a_k\rangle\langle a_k|,
\qquad
B_{\nc}=B-B_{\cl}.
\label{eq:S-Bcl}
\end{equation}
Terms with vanishing probabilities are understood by a limiting prescription.
The classical component is the optimal estimate of $B$ compatible with the
projective measurement of $A$. It commutes with $A$, whereas $B_{\nc}$ contains
the part not resolved as a classical function of that measurement. Hall showed
that
\begin{equation}
\langle B\rangle=\langle B_{\cl}\rangle,
\qquad
\langle B_{\nc}\rangle=0,
\end{equation}
and
\begin{equation}
\langle B^2\rangle
=\langle B_{\cl}^2\rangle+\langle B_{\nc}^2\rangle.
\label{eq:S-variance-decomp}
\end{equation}
For a pure-state unitary trajectory, integrating
Eq.~\eqref{eq:S-hall} gives an exact path-length relation of the type discussed
in Ref.~\cite{Pati:2023seq}.

The Hilbert-Schmidt operator space is itself a Hilbert space. A normalized
operator $|\cO)$ therefore defines a pure state
\begin{equation}
\varrho(t)=|\cO(t))(\cO(t)|.
\end{equation}
Under vectorization,
\begin{equation}
|\cO)\equiv\operatorname{vec}(\cO),
\qquad
\cL\longleftrightarrow
H\otimes\mathbbm 1-\mathbbm 1\otimes H^T,
\label{eq:S-vectorization}
\end{equation}
up to the chosen vectorization convention. Thus Hall's pure-state theorem
applies directly to the normalized operator state.

Let the Krylov basis be generated from a normalized Hermitian seed operator,
with
\begin{equation}
\cL|K_n)=b_{n+1}|K_{n+1})+b_n|K_{n-1}),
\qquad b_0=0,
\label{eq:S-lanczos-hall}
\end{equation}
and expand
\begin{equation}
|\cO(t))
=\sum_{n=0}^{D_{\cO}-1}c_n(t)|K_n),
\qquad
c_n(t)=(-i)^n\phi_n(t),
\label{eq:S-hall-expansion}
\end{equation}
where $\phi_n$ are real and normalized. Define the nondegenerate
operator-space observable
\begin{equation}
A_K=\sum_{n=0}^{D_{\cO}-1}a_n|K_n)(K_n|,
\qquad a_n\neq a_m\quad(n\neq m).
\label{eq:S-AK}
\end{equation}
The measurement probabilities are
\begin{equation}
p_n(t)=(K_n|\varrho(t)|K_n)=\phi_n^2(t).
\label{eq:S-probabilities}
\end{equation}

For $B=\cL$, the numerator in Hall's classical component is
\begin{align}
(K_n|\tfrac12\{\cL,\varrho\}|K_n)
&=\operatorname{Re}\!\left[c_n^*(\cL c)_n\right].
\label{eq:S-classical-numerator}
\end{align}
Since $i\dot c_n=(\cL c)_n$,
\begin{equation}
c_n^*(\cL c)_n=i\phi_n\dot\phi_n,
\end{equation}
which is purely imaginary. Hence
\begin{equation}
\cL_{\cl}^{(K)}=0,
\qquad
\cL_{\nc}^{(K)}=\cL.
\label{eq:S-Lcl-zero}
\end{equation}
This result is relative to the Krylov projective measurement and is not a
basis-independent property of the Liouvillian.

The Liouvillian expectation value vanishes,
\begin{equation}
(\cO(t)|\cL|\cO(t))=(K_0|\cL|K_0)=0,
\end{equation}
and
\begin{equation}
\left(\Delta\cL_{\nc}^{(K)}\right)^2
=(\cO(t)|\cL^2|\cO(t))
=(K_0|\cL^2|K_0)=b_1^2.
\label{eq:S-L-variance}
\end{equation}
Furthermore,
\begin{equation}
\dot p_n=2\phi_n\dot\phi_n.
\end{equation}
At a regular point with $p_n\neq0$,
\begin{equation}
\frac{\dot p_n^2}{p_n}=4\dot\phi_n^2.
\end{equation}
If $\phi_n(t_0)=0$ at an isolated time, the same expression follows by
continuous extension. Therefore,
\begin{equation}
F_K(t)
=\sum_n\frac{\dot p_n^2}{p_n}
=4\sum_n\dot\phi_n^2
=4b_1^2.
\label{eq:S-FK}
\end{equation}
For a pure state translated by $\cL$, the quantum Fisher information is
\cite{BraunsteinCaves:1994}
\begin{equation}
F_Q
=4\left[(\cO|\cL^2|\cO)-(\cO|\cL|\cO)^2\right]
=4b_1^2.
\label{eq:S-FQ}
\end{equation}
Thus
\begin{equation}
F_K(t)=F_Q=4b_1^2.
\label{eq:S-FI-equality}
\end{equation}
At regular points, the fixed Krylov measurement locally attains the QFI. The
measurement is fixed once $H$ and the seed are specified, although its
implementation on the vectorized operator state may be collective or nonlocal
in the original many-body description.

With the Fisher--Rao convention
\begin{equation}
d\ell_{\mathrm{FR}}^2
=\frac14\sum_n\frac{dp_n^2}{p_n},
\end{equation}
we obtain
\begin{equation}
d\ell_{\mathrm{FR}}^2=\sum_n d\phi_n^2.
\end{equation}
Consequently, the actual path length is
\begin{equation}
\ell_{\mathrm{sph}}(0,T)=b_1T,
\end{equation}
and Hall's exact relation becomes
\begin{equation}
T=\frac{\ell_{\mathrm{sph}}(0,T)}
{\Delta\cL_{\nc}^{(K)}}.
\end{equation}
Replacing the actual path length by the shortest endpoint distance produces
the conventional endpoint QSL.

%%%%%%%%%%%%%%%%%%%%%%%%%%%%%%%%%%%%%%%%%%%%%%%%%%%%%%%%%%%%%%%%%%%%%%%%%%%%%%%

\section{Spherical moving frame and local invariants}
\label{supp:geometry}

The real amplitudes obey
\begin{equation}
\dot\phi_n=b_n\phi_{n-1}-b_{n+1}\phi_{n+1}.
\label{eq:S-recursion-geometry}
\end{equation}
With $\dot\Phi=\mathsf A\Phi$, the skew-symmetric generator has
\begin{equation}
\mathsf A_{n,n-1}=b_n,
\qquad
\mathsf A_{n,n+1}=-b_{n+1},
\end{equation}
and therefore
\begin{equation}
\mathsf A\bm e_n=-b_n\bm e_{n-1}+b_{n+1}\bm e_{n+1}.
\end{equation}
Because $\mathsf A^T=-\mathsf A$, the vectors
\begin{equation}
\bm E_n(t)=\e^{t\mathsf A}\bm e_n
\end{equation}
remain orthonormal. With $s=b_1t$, differentiation gives
\begin{subequations}
\begin{align}
\frac{d\bm E_0}{ds}&=\bm E_1,\\
\frac{d\bm E_1}{ds}&=-\bm E_0+\frac{b_2}{b_1}\bm E_2,\\
\frac{d\bm E_n}{ds}
&=-\frac{b_n}{b_1}\bm E_{n-1}
+\frac{b_{n+1}}{b_1}\bm E_{n+1},\qquad n\geq2.
\end{align}
\label{eq:S-moving-frame}
\end{subequations}
Thus the spherical connection coefficients are
\begin{equation}
\kappa_n^{(S)}=\frac{b_{n+1}}{b_1}.
\label{eq:S-spherical-curvatures}
\end{equation}

At $t=0$,
\begin{align}
\dot\Phi(0)&=b_1\bm e_1,\\
\ddot\Phi(0)&=-b_1^2\bm e_0+b_1b_2\bm e_2,\\
\dddot\Phi(0)&=-b_1(b_1^2+b_2^2)\bm e_1
+b_1b_2b_3\bm e_3.
\end{align}
The orthogonality of the flow makes their scalar products time independent. Since
$\dot\Phi\cdot\ddot\Phi=0$, the first ambient Euclidean curvature is
\begin{equation}
\kappa_E
=\frac{\sqrt{\|\dot\Phi\|^2\|\ddot\Phi\|^2
-(\dot\Phi\cdot\ddot\Phi)^2}}
{\|\dot\Phi\|^3}
=\sqrt{1+\frac{b_2^2}{b_1^2}}.
\end{equation}
The first Euclidean torsion, obtained from the Gram determinant of
$\dot\Phi$, $\ddot\Phi$, and $\dddot\Phi$, is
\begin{equation}
|\tau_E|
=\frac{b_2b_3}{b_1\sqrt{b_1^2+b_2^2}}.
\end{equation}
The frame hierarchy is the cleaner formulation in arbitrary dimension because
it does not rely on a three-dimensional cross product.

The relation to moment-based projective geometry follows from
\begin{equation}
\langle\cL^2\rangle=b_1^2,
\qquad
\langle\cL^4\rangle=b_1^4+b_1^2b_2^2.
\end{equation}
Because $\langle\cL\rangle=0$, the Liouville-space curve is a horizontal lift,
and its geodesic curvature obeys
\begin{equation}
\kappa_g^2
=\frac{\langle\cL^4\rangle-\langle\cL^2\rangle^2}
{\langle\cL^2\rangle^2}
=\frac{b_2^2}{b_1^2},
\end{equation}
consistent with earlier quantum-trajectory curvature formulas
\cite{AlsingCafaro:2024}.

For a unit-speed curve on the unit sphere, the covariant acceleration is the
projection of the ambient acceleration onto the tangent space. From
Eq.~\eqref{eq:S-moving-frame},
\begin{equation}
\frac{D\bm E_1}{ds}=\frac{b_2}{b_1}\bm E_2.
\end{equation}
The trajectory is therefore geodesic if and only if $b_2=0$.

Repeated differentiation of Eq.~\eqref{eq:S-recursion-geometry} yields
\begin{align}
\phi_0(t)
&=1-\frac{b_1^2t^2}{2}
+\frac{b_1^2(b_1^2+b_2^2)t^4}{24}+O(t^6),\\
\phi_1(t)
&=b_1t-\frac{b_1(b_1^2+b_2^2)t^3}{6}+O(t^5),\\
\phi_2(t)
&=\frac{b_1b_2}{2}t^2+O(t^4).
\end{align}
Consequently,
\begin{equation}
\phi_0(t)-\cos(b_1t)
=\frac{b_1^2b_2^2}{24}t^4+O(t^6).
\label{eq:S-MT-gap}
\end{equation}
Thus the geodesic curvature controls the leading failure of
Mandelstam--Tamm saturation.

%%%%%%%%%%%%%%%%%%%%%%%%%%%%%%%%%%%%%%%%%%%%%%%%%%%%%%%%%%%%%%%%%%%%%%%%%%%%%%%
\section{Geodesic and nongeodesic qubit examples}
\label{app:qubits}

This appendix provides explicit qubit examples that illustrate the general criteria established in Sec.~\ref{sec:sphere}: the geodesic condition $b_2=0$ and the leading gap from Mandelstam-Tamm saturation controlled by $b_2$, as given in Eq.~\eqref{eq:MT-gap}. These examples are referenced in Sec.~\ref{subsec:MT}.

Consider the Hamiltonian
\begin{equation}
H=\frac{\omega}{2}\sigma_z,
\qquad
|K_0)=\frac{\sigma_x}{\sqrt2},
\label{eq:S-great-circle-H}
\end{equation}
with $\omega>0$. The Krylov basis closes after two sites:
\begin{equation}
|K_1)=\frac{i\sigma_y}{\sqrt2},
\qquad
b_1=\omega,
\qquad
b_2=0.
\label{eq:S-great-circle-coeffs}
\end{equation}
The amplitudes are obtained by solving Eq.~\eqref{eq:amplitude-recursion}:
\begin{equation}
\phi_0(t)=\cos(\omega t),
\qquad
\phi_1(t)=\sin(\omega t),
\label{eq:S-great-circle-amplitudes}
\end{equation}
with all higher amplitudes vanishing. The trajectory is a great circle on the Krylov sphere, with geodesic curvature $\kappa_g=b_2/b_1=0$. Consequently, Eq.~\eqref{eq:return-bound} is saturated while the arc remains minimizing:
\begin{equation}
\phi_0(t)=\cos(\omega t)=\cos(b_1t).
\label{eq:S-great-circle-saturation}
\end{equation}
This explicitly demonstrates that $b_2=0$ is the criterion for Mandelstam--Tamm saturation.

Now consider the Hamiltonian
\begin{equation}
H=\frac{\omega}{2}\sigma_z+\frac{h}{2}\sigma_x,
\qquad
|K_0)=\frac{\sigma_x}{\sqrt2},
\label{eq:S-nongeodesic-H}
\end{equation}
with $\omega,h>0$. A Lanczos basis with positive coefficients is
\begin{equation}
|K_1)=\frac{i\sigma_y}{\sqrt2},
\qquad
|K_2)=-\frac{\sigma_z}{\sqrt2},
\label{eq:S-nongeodesic-basis}
\end{equation}
for which the Lanczos coefficients are
\begin{equation}
b_1=\omega,
\qquad
b_2=h,
\qquad
b_3=0.
\label{eq:S-nongeodesic-coeffs}
\end{equation}
We note that $b_3=0$ terminates the chain at $D_{\mathcal{O}}=3$, so the dynamics is confined to a three-dimensional Krylov subspace.

With $\Omega=\sqrt{\omega^2+h^2}$, the real amplitudes in the convention of Eq.~\eqref{eq:expansion} are
\begin{align}
\phi_0(t)&=\frac{h^2+\omega^2\cos(\Omega t)}{\Omega^2},\\
\phi_1(t)&=\frac{\omega}{\Omega}\sin(\Omega t),\\
\phi_2(t)&=\frac{h\omega}{\Omega^2}\bigl[1-\cos(\Omega t)\bigr].
\label{eq:S-nongeodesic-amplitudes}
\end{align}
The speed remains $\omega=b_1$, but the geodesic curvature is
\begin{equation}
\kappa_g=\frac{b_2}{b_1}=\frac{h}{\omega}.
\label{eq:S-nongeodesic-curvature}
\end{equation}
The path is a non-great circle on $S^2$. The return-amplitude inequality is strict for sufficiently small nonzero times, with the gap given by
\begin{equation}
\phi_0(t)-\cos(\omega t)
=\frac{\omega^2h^2}{24}t^4+O(t^6).
\label{eq:S-nongeodesic-gap}
\end{equation}
This explicitly demonstrates the gap formula ~\eqref{eq:MT-gap}, with $b_1=\omega$ and $b_2=h$. The geodesic curvature $\kappa_g=h/\omega$ controls the first departure from Mandelstam-Tamm saturation.

These two examples illustrate the general criteria established in Sec.~\ref{sec:sphere}: The geodesic case $b_2=0$ saturates the Mandelstam--Tamm bound, while the nongeodesic case $b_2>0$ produces a strict gap whose leading behavior is controlled by $b_2^2$. The speed $b_1$ remains unchanged in both cases, confirming that $b_1$ and the higher coefficients play distinct geometric roles.

%%%%%%%%%%%%%%%%%%%%%%%%%%%%%%%%%%%%%%%%%%%%%%%%%%%%%%%%%%%%%%%%%%%%%%%%%%%%%%%
\section{Individual-amplitude tail bound}
\label{app:tail}

This appendix provides a detailed derivation of the individual-amplitude tail bound presented in Sec.~\ref{sec:index}. The bound follows from elementary properties of the nearest-neighbor Lanczos representation and the Schur row-sum test. We emphasize that this bound is not a consequence of Hall's equality; rather, it is a separate diagnostic for index-space propagation that complements the cumulative nonescape hierarchy of Ref.~\cite{TakahashiDelCampo:2025}.

Let $\{|K_n)\}_{n\ge0}$ denote the orthonormal Krylov basis generated from the initial operator $|K_0)\equiv|\mathcal{O}(0))$, such that the Liouvillian superoperator $\mathcal L$ acts as
\begin{equation}
\mathcal L |K_n)=
b_{n+1}|K_{n+1})+
b_n |K_{n-1}),
\qquad b_0=0,
\label{eq:S-Lanczos}
\end{equation}
with real, non-negative Lanczos coefficients $b_n$. The Heisenberg evolution of the operator can be written as
\begin{equation}
|\mathcal{O}(t)) = e^{-it\mathcal L}|K_0).
\label{eq:S-evolution}
\end{equation}
Projecting onto the Krylov basis defines the Krylov amplitudes $\phi_n(t)$ via
\begin{equation}
|\mathcal{O}(t))=\sum_{n=0}^{\infty}(-i)^n\phi_n(t)|K_n),
\qquad \phi_n(t)\in\mathbb R,
\label{eq:S-expansion-2}
\end{equation}
so that
\begin{equation}
\phi_n(t)= i^{-n}(K_n|e^{-it\mathcal L}|K_0).
\label{eq:S-phi-def}
\end{equation}
Since $|i^{-n}|=1$, this phase plays no role in absolute-value bounds. The amplitudes satisfy the recursion
\begin{equation}
\dot\phi_n=b_n\phi_{n-1}-b_{n+1}\phi_{n+1},
\label{eq:S-recursion}
\end{equation}
with $\phi_n(0)=\delta_{n0}$.

Expanding the exponential in Eq.~\eqref{eq:S-phi-def},
\begin{equation}
e^{-it\mathcal L}
=
\sum_{m=0}^\infty \frac{(-it)^m}{m!}\,\mathcal L^m,
\end{equation}
we obtain
\begin{equation}
\phi_n(t)
=
i^{-n}\sum_{m=0}^\infty \frac{(-it)^m}{m!}
(K_n|\mathcal L^m|K_0).
\label{eq:S-phi-expansion}
\end{equation}

Since the Lanczos representation of $\mathcal L$ is tridiagonal, $\mathcal L^m |K_0)$ has support only on Krylov levels $0 \le n \le m$. This follows by induction. For $m=0$, $\mathcal L^0|K_0)=|K_0)$, which has support on $n=0$. Assume the claim holds for some $m$, so that $\mathcal L^m|K_0)=\sum_{n=0}^m c_n |K_n)$. Acting once more with $\mathcal L$,
\begin{align}
\mathcal L^{m+1}|K_0)
&=\mathcal L\!\left(\sum_{n=0}^m c_n |K_n)\right) \nonumber\\
&=\sum_{n=0}^m c_n\left(b_{n+1}|K_{n+1}) + b_n|K_{n-1})\right),
\label{eq:S-support-induction}
\end{align}
which has support only on $0\le n\le m+1$ (with the understanding that $|K_{-1})=0$). Thus, by induction,
\begin{equation}
(K_n|\mathcal L^m|K_0) = 0,
\qquad\text{for } m<n.
\label{eq:S-support}
\end{equation}
Consequently, Eq.~\eqref{eq:S-phi-expansion} reduces to
\begin{equation}
\phi_n(t)
=
i^{-n}\sum_{m=n}^\infty \frac{(-it)^m}{m!}
(K_n|\mathcal L^m|K_0).
\label{eq:S-phi-truncated}
\end{equation}

To bound the matrix elements in Eq.~\eqref{eq:S-phi-truncated}, we use the operator norm
\begin{equation}
\|\mathcal L\| \equiv \sup_{\|\psi\|=1}\|\mathcal L|\psi)\|.
\end{equation}
For arbitrary normalized states $|\phi)$ and $|\psi)$, submultiplicativity of the operator norm gives
\begin{equation}
|(\phi|\mathcal L^m|\psi)|
\leq \|\mathcal L^m\|
\leq \|\mathcal L\|^m.
\label{eq:S-norm-bound}
\end{equation}

We now bound $\|\mathcal L\|$ using the Lanczos recursion. From Eq.~\eqref{eq:S-Lanczos}, the action of $\mathcal L$ on a basis state $|K_n)$ is
\begin{equation}
\mathcal L|K_n)=b_{n+1}|K_{n+1})+b_n|K_{n-1}),
\label{eq:S-L-action}
\end{equation}
with the convention $b_0=0$. For any state $|\psi)=\sum_n c_n|K_n)$ with $\sum_n |c_n|^2=1$, the norm squared of $\mathcal L|\psi)$ is
\begin{align}
\|\mathcal L|\psi)\|^2
&=\sum_n \left|b_{n+1}c_{n+1}+b_n c_{n-1}\right|^2 \nonumber\\
&\leq \sum_n \left(b_{n+1}|c_{n+1}|+b_n|c_{n-1}|\right)^2.
\label{eq:S-Lnorm-squared}
\end{align}
Using the inequality $(x+y)^2 \leq 2(x^2+y^2)$, we can show that
\begin{equation}
\|\mathcal L|\psi)\|^2 \leq 2\sup_n(b_n^2+b_{n+1}^2)\sum_n |c_n|^2.
\label{eq:S-Lnorm-bound-rough}
\end{equation}
A sharper and more standard bound follows from the Schur row-sum test (equivalently, Gershgorin's circle theorem) applied to the tridiagonal matrix representation of $\mathcal L$. The row sums of the absolute values of the matrix elements are $b_n+b_{n+1}$. Therefore,
\begin{equation}
\|\mathcal L\|
\leq
\LambdaK
\equiv
\sup_{n\ge0} (b_n + b_{n+1}),
\label{eq:S-vop-def}
\end{equation}
provided this supremum is finite. This is the row-sum bound used in the main text.

Combining Eqs.~\eqref{eq:S-norm-bound} and \eqref{eq:S-vop-def}, we obtain
\begin{equation}
|(K_n|\mathcal L^m|K_0)|
\leq
(\LambdaK)^m.
\label{eq:S-matrix-bound}
\end{equation}

Taking absolute values in Eq.~\eqref{eq:S-phi-truncated} and inserting the norm bound \eqref{eq:S-matrix-bound}, we arrive at
\begin{equation}
|\phi_n(t)|
\leq
\sum_{m=n}^\infty \frac{t^m}{m!}\,
|(K_n|\mathcal L^m|K_0)|
\leq
\sum_{m=n}^\infty \frac{(\LambdaK t)^m}{m!}.
\label{eq:S-tail-sum}
\end{equation}

Using the standard bound on the tail of the exponential series,
\begin{equation}
\sum_{m=n}^\infty \frac{x^m}{m!}
\leq
\frac{x^n}{n!}\sum_{r=0}^{\infty}\frac{x^r}{r!}
=
\frac{x^n}{n!}e^x,
\label{eq:S-tail-exponential}
\end{equation}
we obtain
\begin{equation}
|\phi_n(t)|
\leq
\frac{(\LambdaK t)^n}{n!}\,e^{\LambdaK t}.
\label{eq:S-tail-bound}
\end{equation}
This is Eq.~\eqref{eq:amplitude-tail} in the main text.

Applying Stirling's approximation, $n!\sim n^n e^{-n}\sqrt{2\pi n}$ for large $n$, gives the asymptotic form
\begin{equation}
|\phi_n(t)|
\lesssim
\exp\!\left[
-n\log\!\left(\frac{n}{\LambdaK t}\right)
+n
+\LambdaK t
\right],
\label{eq:S-asymptotic}
\end{equation}
which is Eq.~\eqref{eq:decay} in the main text. 
The exponent is negative for
$n\gtrsim c_\ast \LambdaK t$, where
$c_\ast\log c_\ast-c_\ast-1=0$. Thus $|\phi_n(t)|$ decays exponentially once
\begin{equation}
n \gtrsim c_\ast\, \LambdaK t,\;\;\;\;\;\;{\rm with}\;c_\ast\simeq3.59.
\label{eq:S-light-cone}
\end{equation}

This establishes an effective light-cone scale in Krylov index space: For any fixed time $t$, the amplitude at Krylov levels $n$ beyond this scale is exponentially suppressed. The bound is similar to standard Jacobi operator bounds known in the mathematical literature~\cite{Teschl2000}. We emphasize that this bound is not a consequence of Hall's equality; it is a separate diagnostic for index-space propagation that is independent of the geometric interpretation developed in Secs.~\ref{sec:hall-operator} and~\ref{sec:sphere}.

For the semi-infinite uniform chain $b_n=b$ for all $n\ge1$, the row-sum bound gives $\LambdaK =2b$. The exact solution of the amplitude recursion~\eqref{eq:S-recursion} with $\phi_n(0)=\delta_{n0}$ is
\begin{equation}
\phi_n(t)
=J_n(2bt)+J_{n+2}(2bt)
=\frac{n+1}{bt}J_{n+1}(2bt),
\label{eq:S-uniform-exact}
\end{equation}
with the value at $t=0$ defined by continuity. The first equality follows from the standard Bessel function solution to the recurrence; the second is a Bessel identity. For large $n$, the asymptotic behavior of the Bessel functions provides support for $n\lesssim 2bt$, consistent with the light-cone scale $n\sim \LambdaK t=2bt$. The simpler expression $J_n(2bt)$ alone corresponds to a different boundary problem and does not satisfy the semi-infinite initial-value problem at $n=0$. This exact solution demonstrates that the tail bound \eqref{eq:S-tail-bound} captures the correct qualitative behavior, with the ballistic scale $n\sim2bt$.

%%%%%%%%%%%%%%%%%%%%%%%%%%%%%%%%%%%%%%%%%%%%%%%%%%%%%%%%%%%%%%%%%%%%%%%%%%%%%%%

\section{Continuum current and exactly solvable growing chains}
\label{app:growing}

A bond-by-bond estimate based on independent passage times $1/b_n$ fails for
the coherent Krylov recursion and, in particular, misses a factor of 2 for
the solvable chains discussed below. We therefore derive the continuum law
directly from
\begin{equation}
\dot\phi_n = b_n\phi_{n-1} - b_{n+1}\phi_{n+1}.
\label{eq:S-recursion-factor}
\end{equation}
Treat $n$ as continuous and assume a slowly varying, predominantly right-moving
envelope. Expand
\begin{align}
\phi_{n-1}&=\phi-\partial_n\phi+\frac12\partial_n^2\phi+\cdots,\\
\phi_{n+1}&=\phi+\partial_n\phi+\frac12\partial_n^2\phi+\cdots,\\
b_{n+1}&=b+\partial_n b+\cdots.
\end{align}
Substitution into Eq.~\eqref{eq:S-recursion-factor} gives, at leading order,
\begin{equation}
\partial_t\phi
\simeq-2b(n)\partial_n\phi-b'(n)\phi.
\label{eq:S-cont-amplitude}
\end{equation}
For $p=\phi^2$, this becomes a continuity equation,
\begin{equation}
\partial_t p+\partial_n[2b(n)p]\simeq0,
\label{eq:S-continuity}
\end{equation}
which is a first-order hyperbolic conservation law for the probability density $p(n,t)$. The flux is therefore
\begin{equation}
J(n,t)=2b(n)p(n,t).
\label{eq:S-flux}
\end{equation}
For such a conservation law, information propagates along characteristics determined by the derivative of the flux with respect to the conserved density:
\begin{equation}
\frac{dn}{dt}
=
\frac{\partial J}{\partial p}
=
2b(n).
\label{eq:S-characteristics-derivative}
\end{equation}
This is an asymptotic envelope approximation, not a universal rigorous front
bound.

Equation~\eqref{eq:S-characteristics-derivative} yields the characteristic scalings:
\begin{itemize}
\item For constant coefficients $b(n)=b$, the characteristic equation gives $\dot n=2b$, so $n_{\mathrm{typ}}(t)\sim2bt$. This is the ballistic regime.
\item For square-root growth $b(n)=\alpha\sqrt n$, the characteristic equation gives $\dot n=2\alpha\sqrt n$, so $n_{\mathrm{typ}}(t)\sim(\alpha t)^2$.
\item For linear growth $b(n)=\alpha n$, the characteristic equation gives $\dot n=2\alpha n$, so $n_{\mathrm{typ}}(t)\propto e^{2\alpha t}$. This is the exponential regime.
\end{itemize}

These scalings are summarized in Eq.~\eqref{eq:scalings} in the main text. We now verify them with exact solutions.

For the solvable linear-growth family~\cite{Parker:2018yvk,Hornedal:2022pkc}
\begin{equation}
b_n=\alpha\sqrt{n(n-1+\eta)},
\qquad \eta>0,
\label{eq:S-linear-family}
\end{equation}
the Lanczos coefficients grow linearly with $n$ for large $n$: $b_n\sim\alpha n$. The exact real amplitudes are
\begin{equation}
\phi_n(t)
=\sqrt{\frac{(\eta)_n}{n!}}\,
\tanh^n(\alpha t)\,\sech^{\eta}(\alpha t),
\label{eq:S-negative-binomial-amplitude}
\end{equation}
where $(\eta)_n=\Gamma(\eta+n)/\Gamma(\eta)$ is the Pochhammer symbol (the rising factorial). This solution can be verified by direct substitution into the recursion
equation~\eqref{eq:amplitude-recursion}. The derivation proceeds by noting that the generating function for the amplitudes satisfies a simple differential equation, which can be solved in closed form.

The probability distribution is therefore
\begin{equation}
p_n(t)
=\frac{(\eta)_n}{n!}
\bigl[\tanh^2(\alpha t)\bigr]^n
\sech^{2\eta}(\alpha t),
\label{eq:S-negative-binomial}
\end{equation}
which is a negative-binomial distribution with parameters $\eta$ and $\sinh^2(\alpha t)$. Its mean is
\begin{equation}
\mathcal C(t)=\sum_n n p_n(t)
=\eta\sinh^2(\alpha t),
\label{eq:S-mean-linear}
\end{equation}
and, for $\eta>1$, its mode is the integer part of
\begin{equation}
n_{\mathrm{mode}}(t)=(\eta-1)\sinh^2(\alpha t).
\label{eq:S-mode-linear}
\end{equation}
Both scale as $\mathrm{e}^{2\alpha t}$ at late times, confirming the $e^{2\alpha t}$ scaling predicted by the continuum analysis. This is a direct consequence of the factor of 2: The naive estimate using $\Delta t_n\sim1/b_n$ would have given $n\sim e^{\alpha t}$, which is off by a factor of 2 in the exponent.

It is worth noting that this linear growth of Lanczos coefficients appears in many free theories as well, including conformal field theories and free scalar field theories \cite{Dymarsky:2021bjq, Bhattacharjee:2022vlt, Avdoshkin:2022xuw, Camargo:2022rnt, Vasli:2023syq}. The exponential growth of Krylov complexity in these theories is therefore a consequence of the coherent quantum dynamics, not merely the magnitude of the individual hopping amplitudes.

For the square-root chain~\cite{Caputa:2021sib}
\begin{equation}
b_n=\alpha\sqrt n,
\label{eq:S-sqrt-family}
\end{equation}
the exact amplitudes are
\begin{equation}
\phi_n(t)=\mathrm{e}^{-(\alpha t)^2/2}\frac{(\alpha t)^n}{\sqrt{n!}},
\label{eq:S-sqrt-amplitude}
\end{equation}
so the probability distribution is Poissonian,
\begin{equation}
p_n(t)=\mathrm{e}^{-(\alpha t)^2}\frac{(\alpha t)^{2n}}{n!},
\label{eq:S-sqrt-poisson}
\end{equation}
with mean
\begin{equation}
\mathcal C(t)=(\alpha t)^2.
\label{eq:S-sqrt-mean}
\end{equation}
This matches the characteristic scaling $n_{\mathrm{typ}}(t)\sim(\alpha t)^2$ from the continuum analysis.

These exact families confirm the three scalings in Eq.~\eqref{eq:scalings} and demonstrate that index-space transport depends on the full Lanczos sequence, with the characteristic scale determined by the integral of $b(n)$ 
\begin{equation}
n_{\mathrm{typ}}(t) \sim 2\int_0^t ds\, b(n_{\mathrm{typ}}(s)).
\label{eq:S-integral-characteristic}
\end{equation}
This is in stark contrast to the Hall spherical speed, which depends only on the first Lanczos coefficient $b_1$ and is always constant. Accelerated index-space transport, whether ballistic, polynomial, or exponential, coexists with the constant Hall speed on the amplitude sphere, reinforcing the distinction emphasized throughout the main text.

\bibliography{literature-revised}

\end{document}